\documentstyle[epsf]{elsart}

\renewcommand\phi{\varphi}
\renewcommand\epsilon{\varepsilon}

\newcommand\n{\hat{n}}
\newcommand\cc{\hat{c}}

\newcommand\AAeff{|A|^2_{\mathrm {eff}}}
\newcommand\avy[1]{\overline{#1}}

\begin{document}
\begin{frontmatter}
  \title{ 
    Pattern Formation from Defect Chaos ---\\A Theory of
    Chevrons}
  \author{A. G. Rossberg and L. Kramer}
  \address{Physikalisches Institut der Universit{\"a}t Bayreuth,
    D-95440 Bayreuth, Germany}
 \address{(submitted to Physica D 23 December 1996)}
  \date{\today}
  \begin{abstract}
    For over 25 years it is known that the roll structure of
    electroconvection (EC) in the dielectric regime in planarly
    aligned nematic liquid crystals has, after a transition to defect
    chaos, the tendency to form chevron structures.  We show, with the
    help of a coarse-grained model, that this effect can generally be
    expected for systems with spontaneously broken isotropy, that is
    lifted by a small external perturbation. The linearized model as well as 
    a nonlinear extension
    are compared to simulations of a system of coupled
    amplitude equations which generate chevrons out of defect chaos.
    The mechanism of chevron formation is similar to the development
    of Turing patterns in reaction diffusion systems.
  \end{abstract}
\end{frontmatter}

\thispagestyle{myheadings}

\section{Introduction}
\label{sec:introduction}

Understanding defect chaotic states of pattern forming systems is
presently one of the important goals of research in pattern formation
\cite{crohop}.  Important questions  relate to the physical
quantities characterizing properly such a state
\cite{gileme,reh,hilbaer,krila,caci} and to the possibility to
observe and understand transitions between different types of defect
chaos \cite{hilbaer,bjpv}.

An interesting candidate for such a transition is the formation
of chevron patterns occurring in particular in 
the dielectric regime of electroconvection (EC) in
thin layers of planarly aligned nematic liquid crystals
\cite{heihel,orsay2,dvgpp,kaz,books}. Increasing the ac voltage applied 
across the layer (above the "cutoff frequency") one first
observes a (nearly) periodic pattern of (narrow) convection rolls with
wavevector parallel to the director orientation
imposed by surface treatment of the plates forming the upper and lower 
boundaries of the layer. Then defects (dislocations) in the roll pattern
start to appear, their density increasing as the voltage rises. Initially 
they move irregularly and their distribution is homogeneous.
Defects carry a topological charge which is
$+1$ or $-1$ depending on whether a roll ``ends'' or ``begins''
at that point.
After a further increase of the voltage one observes the transition to
chevron patterns, with defects of equal charge ordering along
periodically arranged domains oriented parallel to the original roll
direction.  Simultaneously with the onset of the ordering process the
increase of the number of defects with voltage becomes noticeably
steeper \cite{scheuring}.  Between the chains the convection rolls are
rotated away from their original direction alternatingly to the left
and to the right, which gives the whole structure a ``herringbone''
like appearance.  

In this work it will be argued, that the tendency of a 2-d anisotropic
pattern-forming system to form chevron patterns is a rather general
feature depending in particular on symmetry properties.
As a matter of fact chevron patterns as shown in
Fig.~\ref{fig:chevrons} (for details see Sec.~\ref{sec:numerics}) 
can easily be observed in simulations of the recently proposed 
generic amplitude equations
(\ref{scaled-normal-A},\ref{scaled-normal-phi}) \cite{rohekrpe}.  
This description was inspired by EC in the low-frequency conductive regime
in homeotropically aligned nematics with negative dielectric anisotropy
where one starts from a situation without external anisotropy
(director in the $z$ direction perpendicular to the slab).  Increasing
the voltage one first has a spatially homogeneous Fre\'edericksz
transition where the director bends away from the $z$ direction
thereby singling out spontaneously a direction $\cc$ in the $x$-$y$
plane. The isotropy may also be broken externally by applying a weak magnetic
field parallel to the plain. $\cc$ may then be expressed by the angle
$\phi$ that is enclosed by $\cc$ and the field. At higher voltages
there is an instability to EC with a critical mode corresponding to
rolls with wavevector parallel to the in-plane director $\cc$ (in the
simplest case). A weakly nonlinear description then has to incorporate
the roll mode as well as the undamped mode corresponding to
(infinitesimal) rotations of $\cc$ (the Goldstone mode).  Equations
(\ref{scaled-normal-A},\ref{scaled-normal-phi}) below describe such a
situation.
In fact chevrons have recently been observed in homeotropic EC
\cite{toth}.  Other systems where chevrons were observed are
dielectric EC under oblique boundary conditions \cite{igfre}, EC in
nematic polymers \cite{trublb} and sometimes also in the conductive 
regime of EC in planarly aligned ordinary nematics \cite{hika}.

After introducing our notation and definitions in
Sec.~\ref{sec:formal-setting} we develop a linear and (weakly) nonlinear model
of chevron structures in Secs.~\ref{sec:linear} and
\ref{sec:nonlinear}. In Sec.~\ref{sec:numerics} some predictions of the model
are tested quantitatively against numerical simulations of
Eqs.(\ref{scaled-normal-A},\ref{scaled-normal-phi}). 
Section~\ref{sec:discussion}
discusses experimental and theoretical questions in a more
general framework.

\section{Formal Setting}
\label{sec:formal-setting}

The derivation from symmetry arguments of the coupled amplitude 
equations for systems with spontaneously (and almost spontaneously)
broken isotropy is presented in \cite{rohekrpe}. 
(There the coefficients are also calculated from hydrodynamics
for EC in homeotropically oriented nematics.)
The system should bifurcate supercritically to a roll
pattern with wavevector parallel to $\cc=(\cos\phi,\sin\phi)$ (normal rolls) 
when some
external control parameter $\epsilon$ changes sign from negative to
positive. For small $\epsilon$ and $\phi$ and after
some rescaling the equations take the form
\begin{eqnarray}
  \label{scaled-normal-NW}
  \tau \partial_t A&=&(1+\partial_x^2+(\partial_y - i \phi)^2+i
  \beta_y \phi_{,y}-|A|^2)A
  \label{scaled-normal-A}, \\
  \partial_t
  \phi&=&(K_3\partial_x^2+\partial_y^2-H^2/\epsilon) \phi +
  \label{scaled-normal-phi}\\
  &&\Gamma (-i A^* (\partial_y-i \phi) A + {\it c.c.}). \nonumber 
\end{eqnarray}
In Eq.(\ref{scaled-normal-A}) the derivative operator $\partial_y$
operates only on $A$ and $\phi_{,y}:=\partial \phi/\partial y$. All
coefficients are real, $\tau$ and $K_3$ are positive.  $A$ is the
complex amplitude of the patterning mode.  $A={\it const.}$
corresponds to the most unstable linear mode at $\phi=0$.  Similar to
the usual real Ginzburg-Landau equation 
a factor $\sim \epsilon^{-1/2}$ has been taken out of the physical
length scales, a factor $\sim \epsilon^{-1}$ out of time, and a factor
$\sim \epsilon^{1/2}$ out of $A$ and $\phi$. The spatial extensions of
the plane $L_x$, $L_y$ should be large ($L_x,L_y\gg1$). Whenever needed 
we will impose the convenient periodic boundary conditions. The
coefficient containing $H^2$ describes the (small) external
perturbation of isotropy. (We use this notation since $H$ is typically
a (scaled) magnetic field.) For sufficiently large $H^2$ (or,
equivalently, small $\epsilon$ while $H \ne 0$) one has a
stable band of stationary, spatially periodic solutions of
(\ref{scaled-normal-A},\ref{scaled-normal-phi}). All such solutions
are unstable for $H^2=0$ if $\Gamma$ is negative, which appears to be typical
for nematics.  In simulations one
typically observes a transition to a defect chaotic state when
$h^2:=H^2/(-2 \Gamma \epsilon)$ drops below a critical value of $O(1)$
\cite{rohekrpe}.

Writing the complex amplitude $A$ as $A=|A|\e^{i \theta}$ defines the
phase modulations $\theta$ of the underlying stripe pattern up to
multiples of $2\pi$. This degeneracy plays no roll as long as only
\begin{equation}
  \label{nabla-rho}
  \nabla \theta = {\mathrm{Im}}\left\{\nabla A \over A\right\}
\end{equation}
is considered (and $A \ne 0$). $\nabla \theta$ is the deviation of the
local wavevector from the most unstable wavevector at $\phi=0$ (up to
the rescaling done in Eq.~(\ref{scaled-normal-NW})). The wavevector of the
roll pattern and the director $\cc$ are parallel if $P:=\partial_y
\theta=\phi$.

A topological defect in the stripe pattern corresponds to a simple
zero of the complex amplitude $A$. Its topological charge is $\int
\d{\vec r}\cdot \nabla \theta / 2 \pi$, where the path of integration is a
small loop encircling the defect in the positive sense.

We define defect densities
\begin{equation}
  \label{n-plus-minus}
  n_{\pm}(\vec{r})=\sum_j \delta(\vec{r}-\vec{r}_{\pm,j}),
\end{equation}
where the sum is over all positively or negatively charged defects
at the positions $\vec{r}_{\pm,j}$, respectively, and
$\delta(\cdot)$ is the Dirac $\delta$ function.  The total defect
density is $n:=n_++n_-$ and the topological charge density is $\rho:=n_+-n_-$. 
One has
\begin{equation}
  \label{rho-grad-theta}
  \int_{\partial \Omega}  \nabla \theta(\vec{r})\d\vec{r}
  =2 \pi \int_\Omega \rho(\vec{r}) \d\Omega
\end{equation}
for any area $\Omega$.

From an abstract point of view one can define chevrons as a periodic
modulation of $\rho$, with a wavevector parallel to the $x$ axis
\cite{kaz}.
Hence we will only look at modulations of the defect chaotic state in the $x$
direction and average all equations along $y$,
which will be indicated by an overbar $\avy\cdot := L_y^{-1}\int
\cdot \d y$.  In particular one finds a topological condition
\begin{equation}
  \label{P-vs-rho}
  \partial_x\avy{P}=2 \pi \avy{\rho},
\end{equation}
by differentiating both sides of Eq.~(\ref{rho-grad-theta}) for a
rectangular area $\Omega=(0..\xi) \times (0..L_y)$ with respect to
$\xi$.  Obviously $\avy{P}$ can only change through defect motion.  In
fact, for fixed and (locally) constant $\phi$, where
Eq.~(\ref{scaled-normal-NW}) reduces to the simple Ginzburg-Landau
equation, defects always move such that the (local) wavenumber
approaches the value of highest growth rate \cite{bpk}, which includes
the condition $P=\phi$.

\section{Linear Model}
\label{sec:linear}
A simple model for the coarse-grained dynamics of the defect-chaotic
state helps to understand chevron formation as a linear modulation
instability of homogeneous defect chaos.  Led by symmetry
considerations \cite{rohekrpe} we propose a model equation for the
conservation of topological charge
\begin{equation}
  \label{charge-current}
  \partial_t\avy{\rho}+\partial_x\left[- D \partial_x \avy{\rho} + 
    \sigma (\avy{P} - \avy{\phi} ) \right]=0.
\end{equation}
The diffusion coefficient $D$ and the ``conductivity'' $\sigma$ are
phenomenological constants.  We do not calculate them here and assume
both to be positive. This means that diffusion of defects acts to minimize 
topological charge imbalance and systematic drift acts to reduce 
$|\avy{P}-\avy{\phi}|$.
To get a corresponding equation for $\phi$ we
average Eq.~(\ref{scaled-normal-phi}) over $y$.
\begin{equation}
  \label{avy-normal-phi}
  \partial_t \avy{\phi}=(K_3\partial_x^2+2 \Gamma h^2) \avy{\phi} + 2 \Gamma
    \AAeff (\avy{P}-\avy{\phi}).
\end{equation}
The form of the last term follows from isotropy for small
$\avy{P}-\avy{\phi}$, $\AAeff$ is a suitable constant.
Equations~(\ref{P-vs-rho},\ref{charge-current},\ref{avy-normal-phi})
describe dynamics for small deviations from $\avy{\rho}=0$ on scales
much larger than $n^{-1/2}$.
      
Using this model one can calculate the stability of homogeneous
defect-chaotic solutions against periodic modulations of $\avy{P}$ and
$\avy{\phi}$ in $x$ direction.  After applying $\partial_x$ on
Eq.~(\ref{avy-normal-phi}) and eliminating $\avy{P}$ through
(\ref{P-vs-rho}), Eqs.~(\ref{charge-current},\ref{avy-normal-phi})
take the form of a linearized reaction diffusion system for
$\avy{\rho}$ and $\partial_x \avy{\phi}$.  Accordingly one gets a
homogeneous Hopf bifurcation or a steady state, spatially periodic
(diffusive) Turing instability or a simple non oscillatory,
homogeneous bifurcation as the first instability (see e.g.
\cite{kuramoto,rikra}).

For $K_3/D>1$ and $1<-\Gamma\AAeff/\pi\sigma$, or $K_3/D<1$ and
$(D/K_3)^{1/2}+(K_3/D)^{1/2}< 2(-\Gamma\AAeff/\pi\sigma)^{1/2}$ one
has the Hopf bifurcation at
$h_{\mathrm{Hopf}}^2=\AAeff+\pi\sigma/\Gamma$. The Hopf frequency is
$\omega_{\mathrm{Hopf}}^2=4\pi\sigma(- \Gamma \AAeff -\pi\sigma)$.

\begin{figure}[p]
  \begin{center}
    \leavevmode
    \epsfxsize 8.5 cm
    \epsfbox{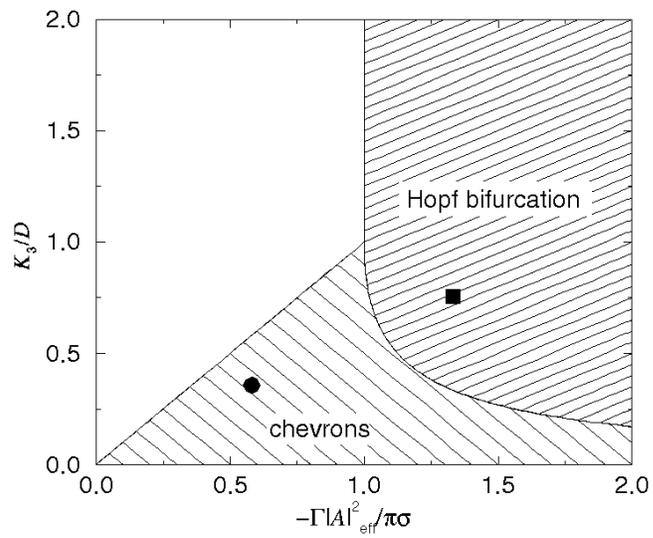}
    \caption{Regions in parameter space where chevron instability and Hopf
      bifurcation respectively are the first to occur when $h^2$ is
      decreased. The circle and the square correspond to the two
      parameter sets analyzed in Section~\protect\ref{sec:numerics}.}
    \label{fig:parameter-space}
  \end{center}
\end{figure}

If $(D/K_3)^{1/2}+(K_3/D)^{1/2}> 2(-\Gamma\AAeff/\pi\sigma)^{1/2}$ and
$K_3/D<- \Gamma \AAeff/\pi\sigma$ the chevron (``Turing'') pattern
with critical wavenumber $k^2_{\mathrm c}=(-4\pi\Gamma\AAeff \sigma /
D K_3)^{1/2}-2\pi\sigma/D$ appears first at $h_{\mathrm c}^2 =
(|A|_{\mathrm{eff} }-(\pi K_3\sigma)^{1/2}(-D\Gamma)^{-1/2})^2$ (see
also Fig.~\ref{fig:parameter-space}).

None of the two instabilities occurs in the remaining case that
$K_3/D>- \Gamma \AAeff/\pi\sigma$ and $1 >-\Gamma\AAeff/\pi\sigma$. At
$h^2=0$ the mode of homogeneous rotation (Goldstone mode) invokes
instability.

According to this model a crucial effect is the diffusive contribution
in Eq.~(\ref{charge-current}). It invokes a wavevector mismatch
$(P-\phi)$ for modulated $P$ and $\phi$ which would otherwise be
leveled out by defect motion.  Due to this mismatch the repulsive
``torque'' on $\phi$ by the roll pattern (described by the term containing
$\Gamma$ in Eq.~(\ref{avy-normal-phi})) can then drive chevron formation.
It should be noted that no particular assumption about the interaction
of single defects was made. By inspection one finds that the model
is robust against additional terms in Eq.~(\ref{charge-current}), as
long as modulations of $P$ are weakened compared to modulations of
$\phi$. For example, a consistent truncation in powers of
$k$ should include a fourth-order diffusive term in
Eq.~(\ref{charge-current}) that has been avoided for the sake of
simplicity.

\section{Nonlinear Model}
\label{sec:nonlinear}

In the final, saturated chevron pattern the defect density
$n$ becomes strongly modulated with half the period of the chevron
pattern. The defects accumulate along ``chains''.  This can be
understood with ideas similar to those describing a p-n junction in a
semiconductor. In the regions along the chains a high density of
positively or negatively charged defects is enforced by the strong
bent ($\partial_x \phi$) of the director. This corresponds to
doping the p and n regions of the semiconductor with charge
carriers. The density of the oppositely charged species is strongly
suppressed due to high recombination probability. As in the depletion
layer of the p-n junction, in the region between the chains the {\em
  total} density of defects $n$ is reduced compared to the ``chain''
region.

The coupling between $\rho$ and $n$ may also provide an important
contribution to the nonlinear saturation of the chevron mode. Assume
that defects are homogeneously created at a constant rate
$n_0/\tau_n$, while they are annihilated at a rate proportional to
$n_+ n_-$, such that for the homogeneous defect chaos one has a 
time-averaged defect density
$\left<n\right>_t=2 n_0$ (In this section we only consider quantities
averaged over $y$ and drop the overbars.) Ignoring
correlations in the nonlinear terms we obtain
\begin{eqnarray}
  \label{n+-dynamik}
  \partial_x j_\pm + \partial_t n_\pm &=& {1\over\tau_n n_0}(n_0^2-n_+
  n_-),
\end{eqnarray}
where $j_{\pm}$ are the defect currents. Since
$n_0^2-n_+n_-=n_0^2-n^2/4+\rho^2/4$ an excitation of the
chevron mode increases the equilibrium number of defects.  One should
expect that with increased defect density $n$ the conductivity $\sigma$
becomes higher (e.g. $\sigma=m n$ with some mobility
constant $m$). This weakens the relative importance of the diffusive
term in Eq.~(\ref{charge-current}) so that the chevron mode saturates. In
order to get a quantitative estimate of this effect we calculated the
weakly nonlinear \cite{croho} steady state solution of
Eqs.~(\ref{avy-normal-phi},\ref{n+-dynamik}), assuming currents of the form
\begin{equation}
  \label{defect-current}
  j_\pm = 
  - D \partial_x n_\pm  \pm m \; n_\pm  (P - \phi),
\end{equation}
and allowing for a modulated value of $\AAeff$ by substituting
\begin{equation}
  \label{AAeff}
  \AAeff \to \AAeff\;(1-c_1 n-c_2 \rho^2-c_3 (P-\phi)^2)
\end{equation}
in Eq.~(\ref{avy-normal-phi}).

For periodic solutions with wavevector $k$ one obtains, up to a
phaseshift,
\begin{eqnarray}
  \label{nl-P}
  P&=&\sqrt{(h_{\mathrm c}^2-h^2) \over G} \sin k x + 
  O(h_{\mathrm c}^2-h^2)^{3/2},\\
  \label{nl-phi}
  \phi&=&(1+{D k^2 \over 4 \pi n_0 m}) P + O(h_{\mathrm c}^2-h^2)^{3/2},\\
  \label{nl-n}
  n&=&2 n_0 + {k^2  (h_{\mathrm c}^2-h^2) \over 16 \pi^2 n_0 G}\cos^2 k x + 
  O(h_{\mathrm c}^2-h^2)^{2},
\end{eqnarray}
with
\begin{eqnarray}
  \label{nl-G}
  G&=&-D k^4 \AAeff\ (64 \pi^2 m^2 n_0^2 (D
  k^2+4 \pi m n_0)^2)^{-1} \times\\
  &&[2 \pi m^3 n_0 + \nonumber\\ 
  &&\quad c_1 (D k^2 m^2 n_0 + 4 \pi m^3 n_0^2)+\nonumber\\
  &&\quad c_2 (4 D k^2 m^2 n_0^2+16 \pi m^3 n_0^3)+\nonumber\\
  &&\quad c_3 (12 \pi D^2 k^2 m n_0 + 3 D^3 k^4]\quad )\nonumber.
\end{eqnarray}
Here $h_{\mathrm c}^2=h_{\mathrm c}^2(k)$ is the value of $h^2$ where
modulations with wavevector $k$ become unstable.

Note the following points: The result is independent of $\tau_n$.
However, this degeneracy can be removed, e.g.  by introducing cross
diffusion terms $-D_2\partial_x n_\mp$ in Eq.~(\ref{defect-current}).
Another consequence of the particular form of
Eqs.~(\ref{n+-dynamik},\ref{defect-current}) is the simple relation
\begin{equation}
  \label{n-vs-rho}
  \tilde n_{2k,0}={|\tilde \rho_{k,0}|^2\over 4 n_0}
\end{equation}
between the Fourier modes of $\rho$ and $n$ [we define the Fourier
transform $\tilde f$ of a function $f$ by $f(x,y)=\sum_{k,l} \tilde
f_{k,l} \exp i(kx+ly)$]. To lowest order one has $n-2n_0\sim\cos^2 k x$, 
so that right between the chains there is no net effect
of the pattern on the value of $n$. Both results should allow a 
simple comparison with experimental chevrons.

\section{Comparison with Simulations}
\label{sec:numerics}

\begin{figure*}[p]
  \begin{center}
    \leavevmode
    \vbox{
      \epsfxsize 8.5 cm
      \epsfbox{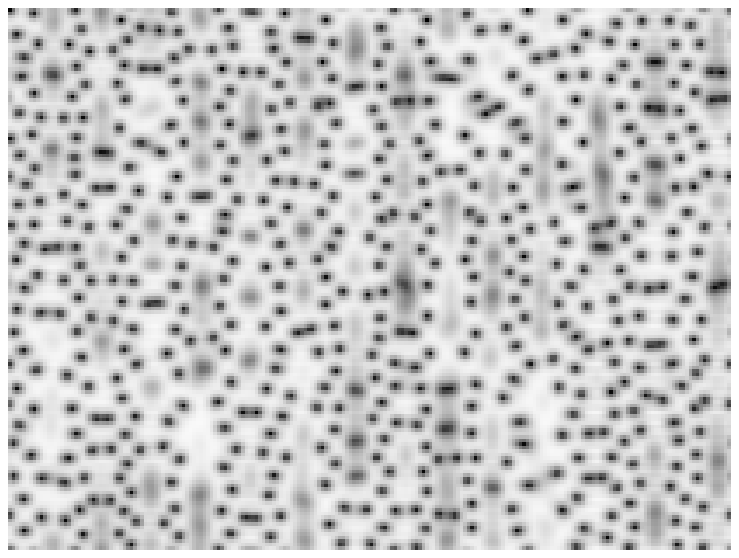}
      \epsfxsize 8.5 cm
      \epsfbox{last.points1.epsi}
      }
    \caption{Snapshot of a simulation of
      Eqs.~(\protect\ref{scaled-normal-A},\protect\ref{scaled-normal-phi})
      with $\tau=4.38738$, $\beta_y=1.07013$, $K_3=0.0564357$,
      $\Gamma=-0.0304092$ in a rectangle of size $159.631 \times 120.25$.  
      The upper image shows $|A|$ encoded in a grey scale, the lower one
      shows the positions and polarity of defects as detected by the algorithm
      used throughout this work.}
    \label{fig:chevrons}
  \end{center}
\end{figure*}

As pointed out in the introduction, chevron patterns are easily
generated by simulating our ``microscopic equations''
(\ref{scaled-normal-A},\ref{scaled-normal-phi}).
Figures~\ref{fig:chevrons} and \ref{fig:h005} show snapshots of the
steady state for different sets of parameters.

For the semi-quantitative test of our model we used a pseudo-spectral
algorithm on a $128 \times 128$ grid with spatial extensions
$L_x=159.63$, $L_y=60.125$ and periodic boundary conditions.

The set of parameters
in~(\ref{scaled-normal-A},\ref{scaled-normal-phi}) was chosen in such
a way that the chevron bifurcation sets in with a small wavevector.
Moreover we made sure, that
the oscillatory mode is absent.  These conditions are satisfied by
keeping
\begin{equation}
  \label{parameter}
  \begin{array}{rclcrcl}
    \tau&=&1.53558,&\em&\beta_y&=&1.07013,\\
     K_3&=&0.180594,&\em& \Gamma&=&-0.0304092
  \end{array}
\end{equation}
fixed, while taking $h^2$ as a control parameter. We remind that
the chevron amplitude increases by {\em lowering} $h^2$. The condition
of small $k_{\mathrm c}$ implies also $h^2_{\mathrm c}(k_{\mathrm c})$
to be small and consequently the supercritical range is rather small,
too.

\begin{figure*}[p]
  \begin{center}
    \leavevmode
    \epsfxsize 8.5 cm
    \epsfbox{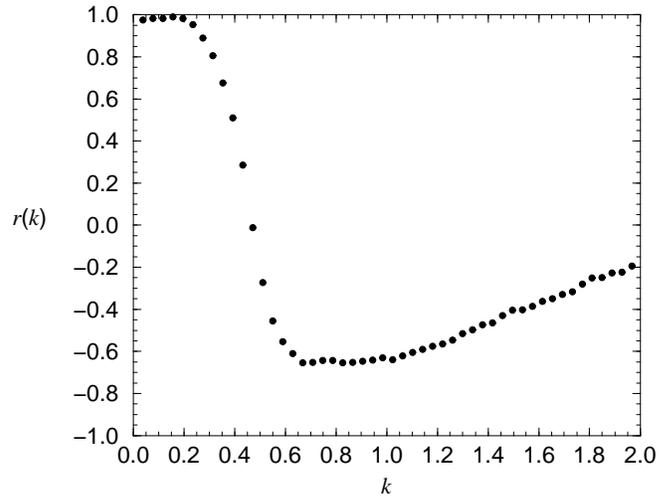}
    \caption{Correlation between Fourier modes (see text).}
    \label{fig:correlation}
  \end{center}
\end{figure*}

The most basic assumption of the model is that, to linear order, the
dynamics of $\avy{P}$ and $\avy{\phi}$ may be isolated from the
remaining degrees of freedom and can be described by a set of linear
PDEs, even though the full dynamics of $A$ is strongly nonlinear. The
coupling to the other degrees of freedom should be describable by
adding noise.  Close to the chevron instability, where one of the
branches of eigenvectors of these linear PDEs is strongly excited by
the noise, there should be a strong correlation between the Fourier
modes of $P$ and $\phi$.  Figure~\ref{fig:correlation} shows the
correlation coefficient $r(k)={\mathrm{Re}}\langle \tilde P_{k,0}
\tilde\phi_{k,0}^*\rangle_t/(\langle |\tilde P_{k,0}|^2 \rangle_t\langle
\tilde|\phi_{k,0}|^2\rangle_t)^{1/2}$ for $h^2=0.025$. Clearly there 
is essentially perfect correlation for $k < 0.24$. The sharp drop of
$r(k)$ at $k\approx0.3$ may be partly due to a change of the relative
sign of $P_{k,0}$ and $\phi_{k,0}$ in the linear mode. 

\begin{figure*}[p]
  \begin{center}
    \leavevmode
    \epsfxsize 8.5 cm
    \epsfbox{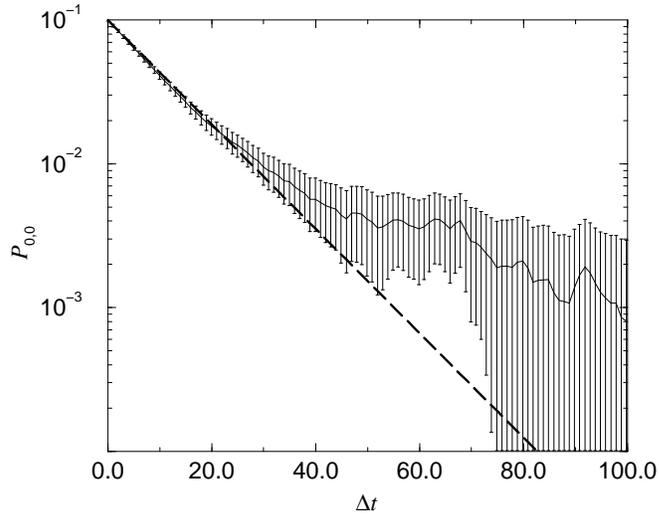}
    \caption{Relaxation of $\tilde P_{0,0}$ after a jump in
      $\tilde\phi_{0,0}$ from $\phi_0=0.1$ to $0$, averaged over 59
      runs. The dashed line is a least square fit.}
    \label{fig:relaxation}
  \end{center}
\end{figure*}

In order to test the usefulness of the particular
form~(\ref{charge-current},\ref{avy-normal-phi}) of
the PDEs we first developed methods to ``measure'' $\sigma$ and $D$ in
simulations. To find $\sigma$, the Fourier mode $\tilde
\phi_{0,0}$ of $\phi$ is fixed at some finite value $\phi_0$, while
all other modes evolve according to Eq.~(\ref{scaled-normal-phi}). This
was implemented by resetting $\tilde \phi_{0,0}$ to $\phi_0$ after each time 
step.  The simulation is run until a steady state is reached. Then
$\phi_0$ is set equal to zero. The relaxation rate of
the global average of $P$ equals $2\pi\sigma$. Figure~\ref{fig:relaxation}
shows the average value of $P$ as a function of the time $\Delta t$
after switching $\phi_0$ to $0$. Within the accuracy obtained,
the decay is exponential.

Similarly, in order to measure $D$, a single pair of long wavelength
Fourier modes $\tilde \phi_{k_0,0}$, $\tilde \phi_{-k_0,0}$ of $\phi$
was held fixed at a finite value $\phi_0$.  $D/\sigma$  can be 
calculated from the steady state average value of $\tilde P_{k_0,0}$ by
setting $\partial_t \avy{\rho}=0$ in Eq.~(\ref{charge-current}).
The values for $\sigma$ and $D$ obtained for
different values of $\phi_0$ and $k_0$ were consistent.
As a simple estimate for $\AAeff$ and $n_0$ we took the spatial and
temporal average of $|A|^2$ and $n/2$.

The numerically calculated coefficients at $h^2=0.025$ are
\begin{equation}
  \label{num-coeffs}
  \begin{array}{rclcrcl}
    \AAeff&=&0.7996(4),&\em& n_0&=&0.00509(10), \\
    D/\sigma&=&38.0(1.3),&\em& \sigma&=&0.01330(73).
  \end{array}
\end{equation}
Error estimates contain only stochastic contributions. From the linear
model one then calculates the threshold $h^2_{\mathrm c}=0.037(5)$ and
the critical wavenumber $k_{\mathrm c}=0.214(5)$.

\begin{figure*}[p]
  \begin{center}
    \leavevmode
    \epsfxsize 8.5 cm
    \epsfbox{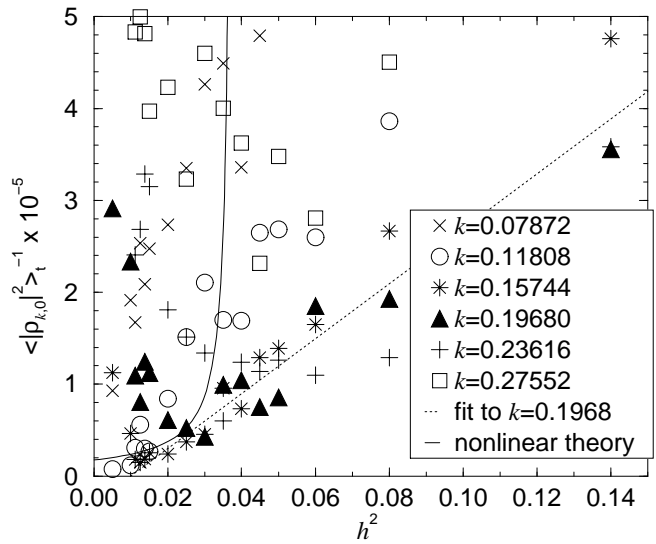}
    \caption{Test of the linear and nonlinear model, see text.}
    \label{fig:threshold}
  \end{center}
\end{figure*}

\begin{figure*}[p]
  \begin{center}
    \leavevmode
    \vbox{
      \epsfxsize 8.5 cm
      \epsfbox{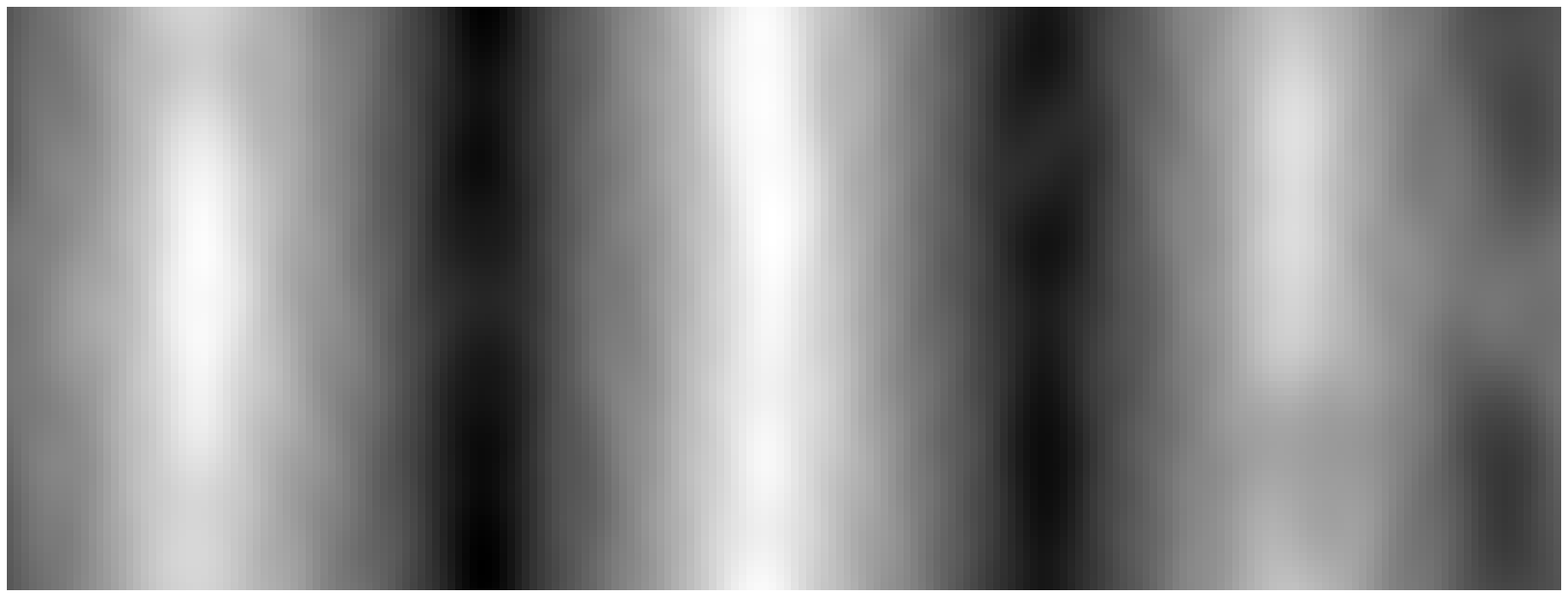}
      \epsfxsize 8.5 cm
      \epsfbox{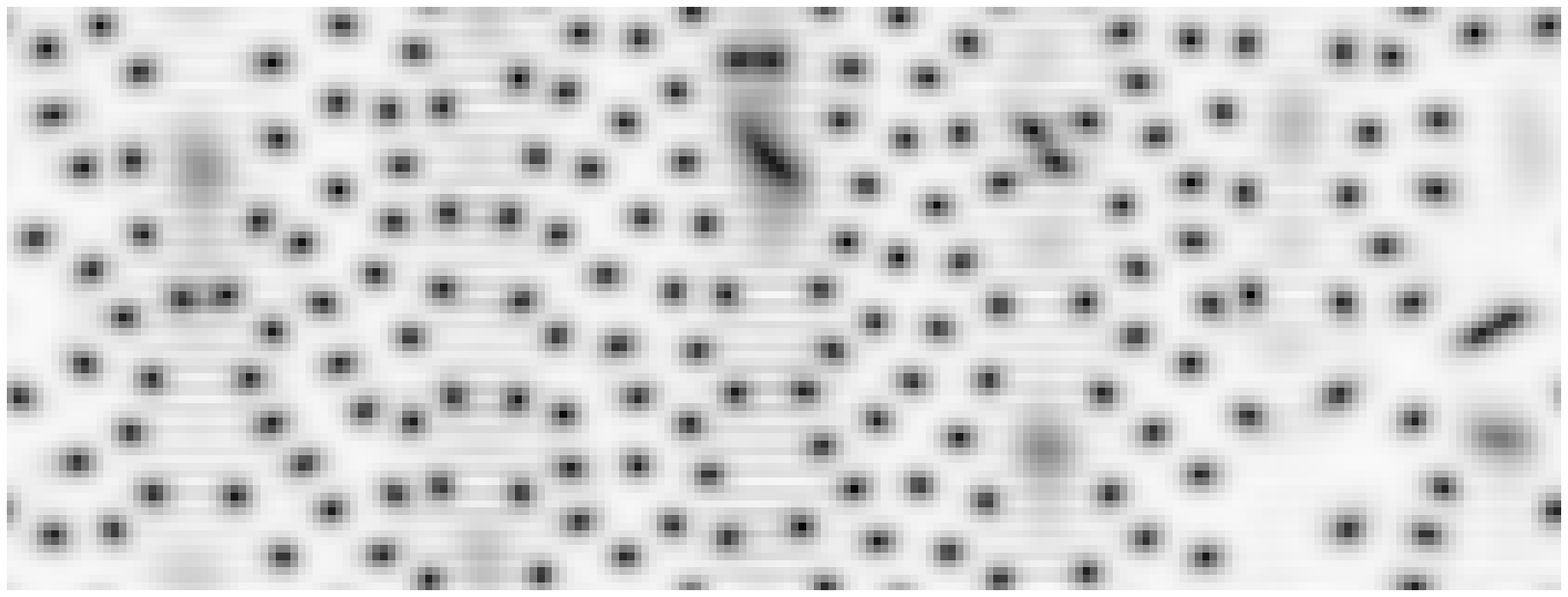}
      \epsfxsize 8.5 cm
      \epsfbox{last.points.epsi}
    }
    \caption{Snapshot of the steady state corresponding to
    $h^2=0.005$ in Fig.~\protect\ref{fig:threshold}. Top: values of
    $\phi$ ($-2.33\ldots2.30$) encoded on a gray scale; center: $|A|$;
    bottom: defect positions and polarities.}
    \label{fig:h005}
  \end{center}
\end{figure*}

Critical slowing down and strong noise make a precise determination of
the threshold in simulations difficult.  Since we expect $h_{\mathrm
  c}^2$ to be close to zero, where the Goldstone mode becomes
unstable, the straight forward method of extrapolating the
supercritical modulation amplitude to zero cannot be applied. Instead
we looked at the subcritical excitation of linear modes by noise.
Without nonlinear interaction one expects a relation of the type
$\left<|\tilde \rho_{k,0}|^2\right>_t^{-1}\sim -s_k(h^2)$, where
$s_k(h^2)$ is the growth rate of the linear mode at wavevector $k$.
Figure~\ref{fig:threshold} shows some numerical results for
$\left<|\tilde \rho_{k,0}|^2\right>_t^{-1}$ at various values of $k$.
The dotted line is a linear fit to $k=0.19680$, which implies a
threshold value of $h_{\mathrm{c,num}}^2 \approx 0.01$. At
$h^2_{\mathrm{c,num}}$, however, the $k=0.19680$ mode itself is already
strongly suppressed by nonlinear interaction with other modes. We
believe that the discrepancy between $h_{\mathrm c}$ and
$h_{\mathrm{c,num}}$ is to a large part an effect of the combination
of nonlinearity and strong noise. For further discussion see next
section. Figure~\ref{fig:h005} shows the fully excited chevron pattern
at $h^2=0.005$.

\begin{figure*}[p]
  \begin{center}
    \leavevmode
    \epsfxsize 8.5 cm
    \epsfbox{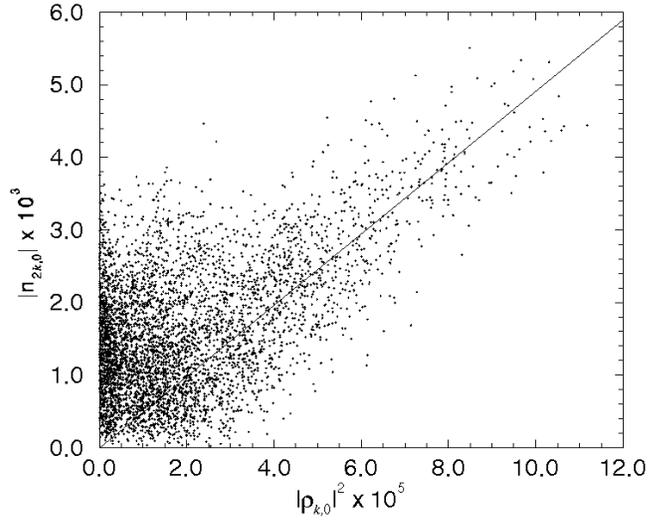}
    \caption{Samples of pairs $(|\tilde n_{2k,0}|,|\tilde
      \rho_{k,0}|^2)$ at  $k=0.19680$ and $h^2=0.025$ from numerical
      simulations and the relation
      given by (\protect\ref{n-vs-rho}) (solid line).}
    \label{fig:rho2-n}
  \end{center}
\end{figure*}

We tested two aspects of the nonlinear model. Firstly, we find
Eq.~(\ref{n-vs-rho}) to be well satisfied, as long as only
one mode is active. In Fig.~\ref{fig:rho2-n} the distribution
of pairs $(|\tilde n_{2k,0}|,|\tilde \rho_{k,0}|^2)$ and the
theoretical line for $k=0.19680$ and $h^2=0.025$ are shown. While $|\tilde
\rho_{k,0}|^2$ is large the other modes are suppressed nonlinearly and
the single mode approximation made in Eq.~(\ref{n-vs-rho}) is
legitimate whereas when $|\tilde \rho_{k,0}|^2$ is small the influence of
competing modes becomes noticeable.

Secondly, we tested the prediction of Eq.~(\ref{nl-P}) with
(\ref{nl-G}) and (\ref{P-vs-rho}) for the chevron amplitude.
Due to the intricate situation at threshold our simple result 
cannot give more than an order of magnitude estimate.
The solid line plotted in Fig.~\ref{fig:threshold}, shows
the prediction of formula~(\ref{fig:threshold}).  It was calculated
using coefficients as for the linear model with $k=k_{\mathrm c}$,
$m=\sigma/2 n_0=1.31(8)$ and $c_1,c_2,c_3=0$ without any fitting. It
is easily seen that in our case the terms containing $c_1,c_2,c_3$
give only small corrections as long as $c_1,c_2,c_3=O(1)$.

\begin{figure*}[p]
  \begin{center}
    \leavevmode
    \epsfxsize 8.5 cm
    \epsfbox{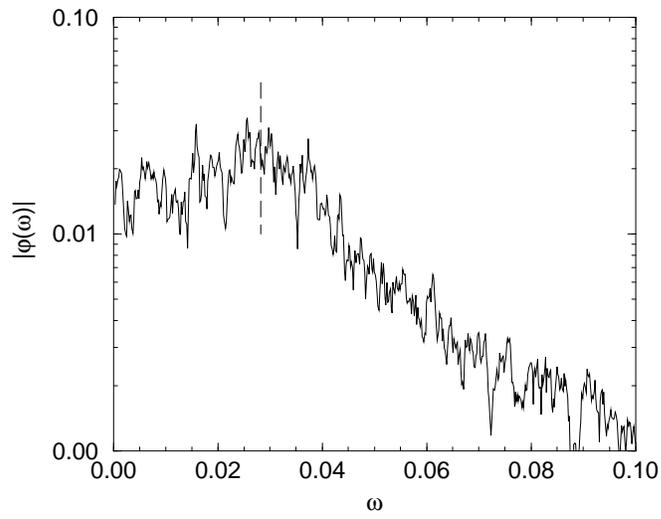}
    \caption{Fourier transform of a simulation of
      Eqs.~(\protect\ref{scaled-normal-A},\protect\ref{scaled-normal-phi})
      in the oscillatory regime
      and the frequency of the most unstable mode of
Eqs.~(\protect\ref{P-vs-rho}--\protect\ref{avy-normal-phi})
      (dashed line). The parameters are given in the text.}
    \label{fig:frequency}
  \end{center}
\end{figure*}

The Hopf bifurcation predicted by our model can actually be found in 
simulations of Eqs.~(\ref{scaled-normal-A},\ref{scaled-normal-phi}). 
For example, at $\tau=
2.19369$, $\beta_y=1.07013$, $K_3=0.338614$, $\Gamma= -0.0608184$ and
$h^2=0.1$ one gets $\AAeff=0.7409(5)$, $\sigma = 0.01075(36)$ and
$D/\sigma = 44.4(2.2)$. From this the Hopf threshold is found to be
$h^2_{\mathrm{Hopf}}=0.186(19)$. The frequency of the $k=0$ eigenmode is
$\omega^2=-4\pi \Gamma \AAeff \sigma-[\Gamma
(\AAeff-h^2)-\pi\sigma]^2=(0.0282(7))^2$ which is in good agreement
with the maximum of the Fourier transform of the simulated time series
of $\tilde \phi_{0,0}$ shown in Fig.~\ref{fig:frequency}.

\section{Discussion}
\label{sec:discussion}

As mentioned in the introduction the most striking case of
experimental chevron formation occurs in the dielectric regime of
electroconvection in planarly oriented nematics. At first glance one
would not expect this system to fall into the class described by the
theory, Eqs.~(\ref{scaled-normal-A},\ref{scaled-normal-phi}), because
here the horizontal orientation of the nematic director $\n$, which
would represent the isotropy breaking degree of freedom, is fixed by
the boundaries which have been prepared to align the director in one
direction. However, in the dielectric regime the wavelength of the
convection pattern is essentially independent of, and usually much
smaller than the thickness of the layer $d$ \cite{ehcbook}.  Then the
boundaries can be reduced to a perturbative effect that may be
described by the term $H^2 \phi$ in Eq.~(\ref{scaled-normal-phi}).

It is not easy to verify the validity of our model for planar
dielectric EC. The most important prediction is the relation between
the local wavevector and the director. A straight forward polarimetric
measurement of the director making use of the birefringence of the material,
however, is not possible, since the
polarization axis of light passing the probe follows the director essentially
adiabatically. The polarization axis is therefore determined by the boundaries.
Some evidence for our model can be taken from the fact, that in the
regime of strongly developed chevrons, where the local wavevector is
turned by nearly $\pi/2$, new structures form inside the nematic, which
superimpose with the chevron pattern.  They allow an interpretation as
disclination lines resulting from the fact that in the midplane the
degenerated directions $\n$ and $-\n$ reconnect, which cannot occur
at the boundaries \cite{igfre}.

Defects in experimental chevrons can align much better along chains
than has been found in the simulations. This is presumably because
higher-order terms and non-adiabatic effects that couple the defect cores
to the underlying roll structure are not included in the amplitude equations.
When experimental chevron patterns are strongly excited it is usually
observed, that defects start to move along  the chains,
alternatingly `up' and `down', i.e. there is a correlation between
$\partial_x \phi$ and the $y$ component of the velocity of a defect
(ignoring its charge). This can not be observed in simulations of
Eqs.~(\ref{scaled-normal-A},\ref{scaled-normal-phi}). In fact, it is
forbidden by the accidental symmetry $\phi\to -\phi$, $A\to A^*$ of
Eqs.~(\ref{scaled-normal-A},\ref{scaled-normal-phi}). Inclusion of
higher order terms would break this symmetry. In contrast the
invariance under $\phi\to-\phi$, $y\to-y$, which follows from inversion
symmetry \cite{rohekrpe}, remains.

We wish to emphasize that chevrons cannot appear directly from the
homogeneous (basic) state via a stationary supercritical bifurcation.
Although this is consistent with most experiments there is some
evidence that at sufficiently high frequencies chevrons can be observed
directly at onset of the dielectric instability \cite{kaz}.  This would
mean that either the primary bifurcation becomes subcritical, which is
not expected from theory, or there is a (``hidden'') bifurcation leading
to an inhomogeneous state already below the dielectric instability.
According to some older measurements the possibility of a first
transition to a state with ``wide domains'' (with width $\sim d$)
appears to exist \cite{ridu,nalugiu,bbgt,trublb}, but this phenomenon
has not been cleared up. In any case this cannot be the general
explanation.

It has been speculated that chevrons can be understood as a kind of 
interference effect between the two most unstable dielectric modes, 
which differ in their $z \to -z$ symmetry. Because of the character of the
dielectric rolls (wavelength much smaller than $d$) the growth rates of these
modes differ very little (as is the case for the higher z modes)
\cite{kaz}. The model has not been worked out, and can, in our view,
not explain chevrons.

It might be interesting to approach the chevron transition from the
point of view of phase transition theory. We are dealing with a
breaking of a continuous symmetry in two dimensions of the
same symmetry class as the $x$-$y$ model. Therefore the transition is,
at least when nonadiabatic effects are neglected,
possibly of the Kosterlitz-Thouless type \cite{kothou}. The model
given here represents a simplified, mean-field type description.
Consequently it is not surprising that the transition is delayed and 
difficult to pin down in the
full ``microscopic'' theory described by
Eqs.~(\ref{scaled-normal-A},\ref{scaled-normal-phi}).

Finally we wish to point out that chevron-like structures have also been 
found in simulations of the anisotropic version of the well-known
complex Ginzburg-Landau equation \cite{advmat}. The mechanism operative 
here is presently under investigation.

\ack{
We wish to thank W.~Pesch for useful discussions and the experimental 
groups in Budapest (A.~Buka and P.~Toth), Leipzig (H.~Amm and R.~Stannarius)
and Bayreuth (C.~Haite, J.~Peinke, and M.~Scheuring) 
for sharing their results with us previous to publication.
Financial support by Volkswagen Stiftung and Deutsche Forschungsgemeinschaft
is gratefully acknowledged.
}

\bibliographystyle{prsty}
\bibliography{/home/btp424/bib/bibview}

\end{document}